\documentstyle[preprint,aps,psfig]{revtex}
\input{psfig}
\def\be{\begin{equation}}
\def\ee{\end{equation}}
\def\bea{\begin{eqnarray}}
\def\eea{\end{eqnarray}}

\begin{document}

\title{Anisotropic flow in ultra-relativistic heavy ion collisions}

\author{Marcus Bleicher${}^{a \xi}$\thanks{E-mail: 
bleicher@nta2.lbl.gov} and 
Horst St\"ocker${}^{b}$, 
}

\address{${}^a$ Nuclear Science Division,
Lawrence Berkeley Laboratory, Berkeley, CA 94720, U.S.A.}

\address{${}^b$Institut f\"ur
Theoretische Physik,  Goethe-Universit\"at,
60054 Frankfurt am Main, Germany}

\footnotetext{${}^\xi$ Feodor Lynen Fellow of the Alexander v. 
Humboldt Foundation}


\maketitle

\begin{abstract}
A sign reversal of the directed flow parameter $v_1$ in the central 
rapidity region in Au+Au collisions at $\sqrt s =200$~AGeV is predicted.
This anti-flow is shown to be linked to the expansion of the
hot matter created.
In line with this observation the predicted elliptic flow 
parameter $v_2$ of various particle species is linked to 
the mean free path of these particles.\\
\underline{LBNL-preprint: LBNL-46096}
\end{abstract}


The exploration of the transverse collective flow is
the earliest predicted observable to probe heated and 
compressed nuclear matter \cite{scheid68a}.
Its sensitivity to the equation of state (EoS) might be used to search 
for abnormal matter states and phase 
transitions \cite{hofmann76a,stoecker86a,Stocker:1979mj}.

Until now, the study of directed and anisotropic flow in high 
energy nuclear collisions is attracting large attention from both
experimentalists and theorists~\cite{Csernai:1982ff,H-G,E877,liu97,posk98}. 
Flow in general is sensitive to the equation of
state~\cite{Csernai:1982ff,H-G,sorge99,soff,bravina} which governs 
the evolution of the system 
created in the nuclear collision. Elliptical 
flow~\cite{sorge99,olli92,lshur,risc96,sorge97,heisel,csernai,brachmann,zhang99}
(i.e. squeeze-out except for a reversed sign in the observable)
is especially sensitive to the early time scales of the reaction. It might 
serve as a keyhole to the (non-)equilibrium dynamics of the 
strongly interacting matter even before hadronization.

In fluid dynamics, the transverse collective flow is intimately 
connected to the pressure $P(\rho, S)$  (which in turn depends on 
the density $\rho$ and the entropy $S$) of the matter  in the reaction
zone \cite{stoecker81a}:
\begin{equation}
\label{pxeqn}   
{\bf p}_x \,=\, \int_t \int_A P(\rho,S) \, {\rm d}{\bf A} \, {\rm d}t\,.
\end{equation}
Here d{\bf A} represents the surface element between the participant and
spectator matters and the total pressure is the sum of the
potential pressure and the kinetic pressure. Thus, the transverse 
collective flow depends directly on the equation of state, $P(\rho,S)$.

In the particle cascade picture employed in transport theoretical approaches, 
the collective expansion of the system created during a heavy-ion 
collision also implies space-momentum correlations in particle distributions
at freeze-out. However, it is unclear how this collective motion 
is connected to the hydrodynamical flow behaviour of the matter, since 
one deals with a system out of equilibrium.  
Nevertheless, a strong space-momentum correlation in the particle emission
pattern persists: This means that - on average -  particles 
created on the left  side of the system move in the 
left direction and particles created 
on the right side move in the right direction.
We will show that the rapidity dependence of directed flow
of nucleons and pions can be used to address this space-momentum correlation 
experimentally. Therefore, it can yield an 
undisturbed view into the properties of the expanding matter \cite{wiggle}.

\begin{figure}[t]
\vskip 0mm
\vspace{1.0cm}
\centerline{\psfig{figure=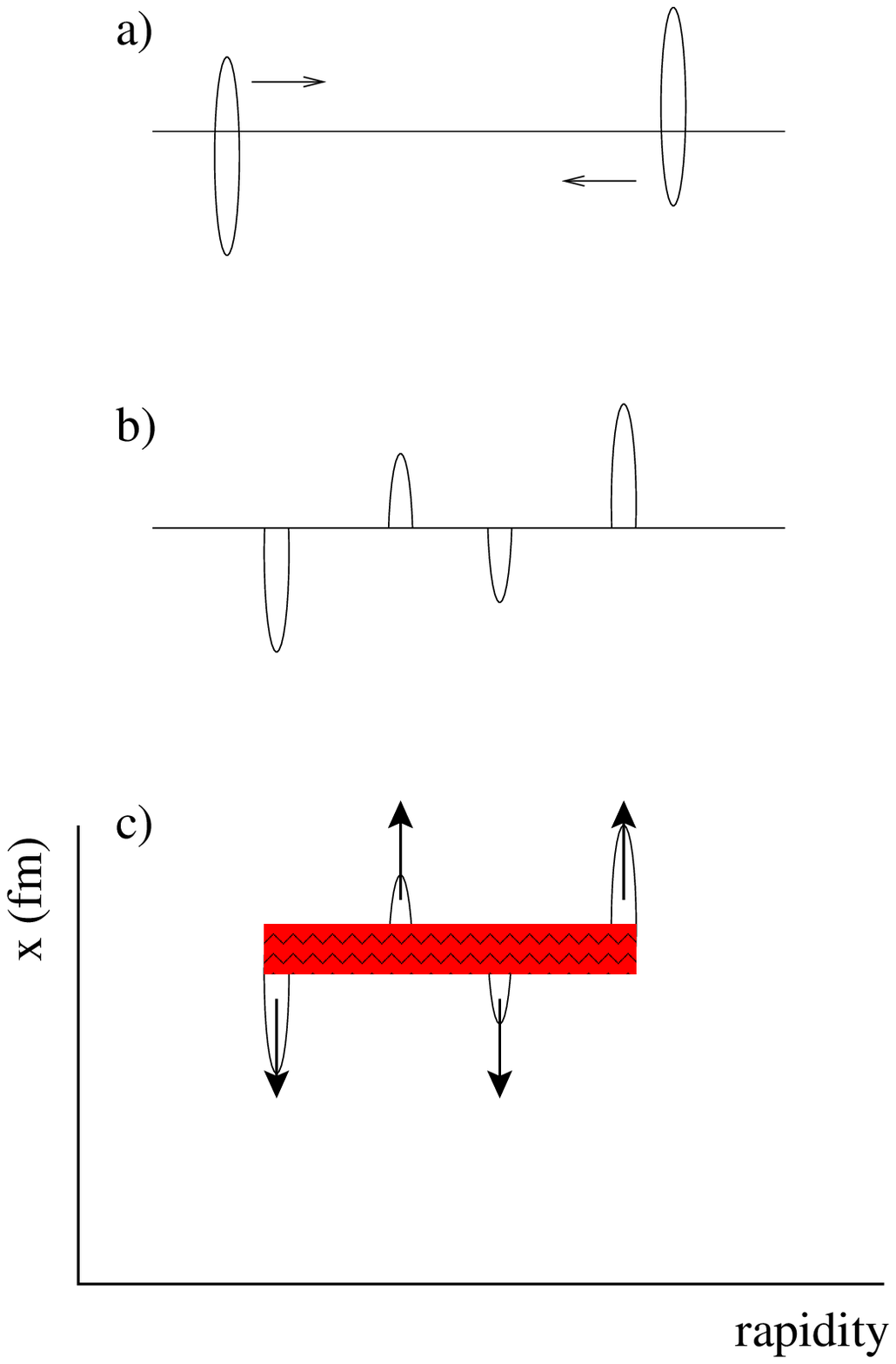,width=10cm,clip=}}
\vskip 2mm
\caption{A schematic sketch of a semi-central heavy-ion collision
in progressing time (a) to (c) and how the excited matter (grey area in (c))
leads to the antiflow of nucleons at central rapidities.
In these figures the vertical axis is the
coordinate along the impact parameter direction and the horizontal axis 
is the rapidity axis. 
\label{negativeflow}}
\end{figure}

\begin{figure}[t]
\vskip 0mm
\vspace{1.0cm}
\centerline{\psfig{figure=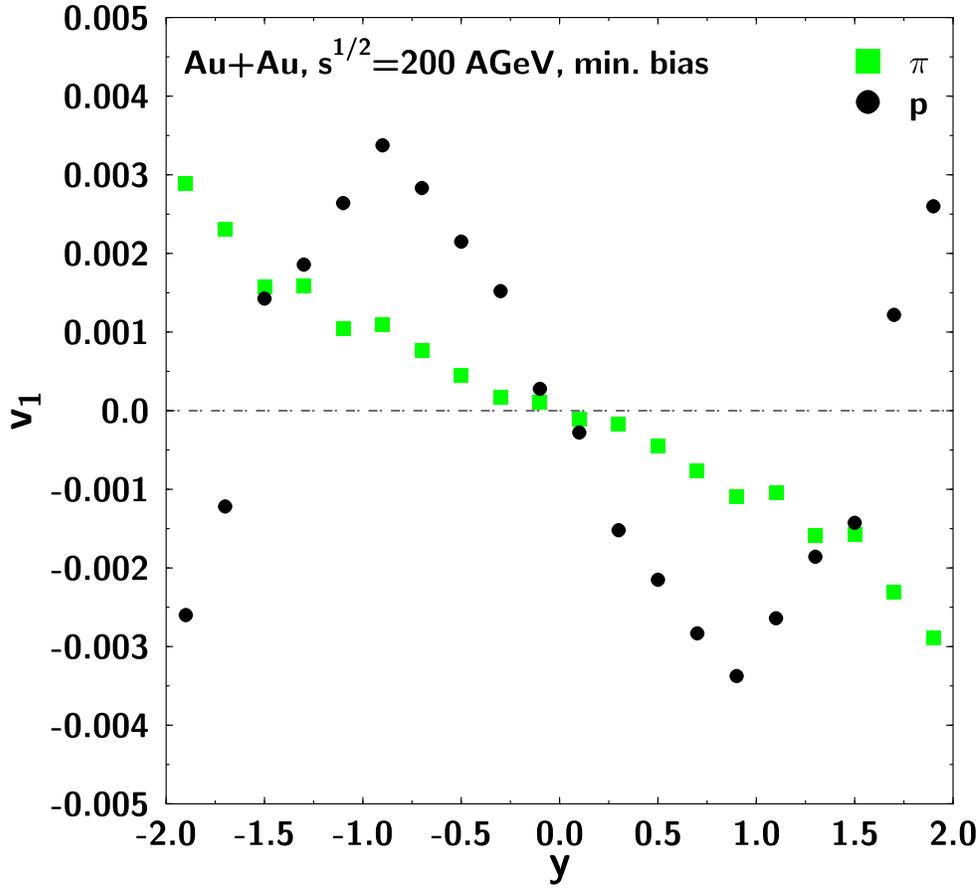,width=15cm,clip=}}
\vskip 2mm
\caption{Antiflow of protons in the central rapidity region as observed 
in the directed flow parameter $v_1$ at central rapidities in
Au+Au reactions at $\sqrt s =200$~AGeV, min.bias.
\label{v1y_22}}
\end{figure}

A sketch of a semi-central heavy-ion collision
is shown in Fig.~\ref{negativeflow}, from before the collision (a) 
to the resulting distributions of $\langle x \rangle$ and 
$\langle p_x \rangle$ shown in (c).
In Fig.~\ref{negativeflow}a the projectile and target are shown before 
the collision in rapidity (horizontal axis) and coordinate 
space (vertical axis, labeled x) - the off-set between the nuclei 
is given by the impact parameter.  
In Fig.~\ref{negativeflow}b the overlap region of the collision is shifted due
to interactions towards the midrapidity region. The spectators are only
slightly affected by the collisions and remain essentially at target and
projectile rapidity. 
The undressing of the participating nucleons results in a rapidity shift 
of the nucleons and leads to the creation of hot
matter along the beam axis (shown in Fig.~\ref{negativeflow}c as grey area
around the horizontal axis). The subsequent expansion of this highly excited
vacuum leads to a positive space-momentum correlation and the nucleons 
are pushed away from the beam axis. Thus, nucleons around central rapidities 
sitting at negative $x$ in coordinate space 
will receive negative $\langle p_x \rangle$-push, while those at positive 
$x$ acquire a positive $\langle p_x \rangle$. 
This results in a wiggle structure \cite{wiggle} in the rapidity 
dependences of
$\langle p_x \rangle$ (or $v_1$, respectively) 
which is depicted in Fig.~\ref{v1y_22} (and will be discussed in detail below).

The shape of the wiggle, the magnitude of $v_1$ and the rapidity 
range, depend on the pressure exerted by the partons and color fields 
creating the space-momentum 
correlation and therefore on the equation of state of the hadronizing 
matter \cite{sorge99a}.  Indeed, comparing the UrQMD model predictions 
to RQMD calculations \cite{wiggle} which include a rope mechanism to mimic
parton interactions in the early stage, the non-interacting strings employed in
the UrQMD approach result in smaller space-momentum 
correlations and thus in a smaller 'anti-flow` around midrapidity.

\begin{figure}[t]
\vskip 0mm
\vspace{1.0cm}
\centerline{\psfig{figure=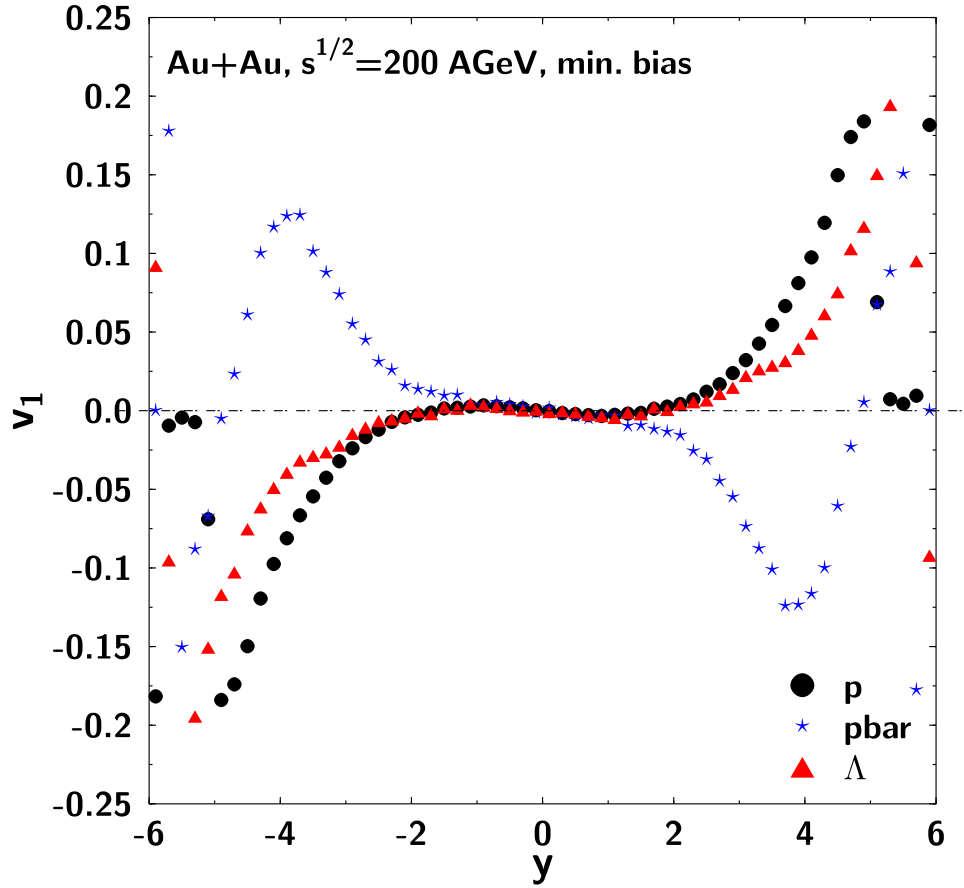,width=15cm,clip=}}
\vskip 2mm
\caption{Directed flow parameter $v_1$ of protons, lambdas and anti-protons 
as a  function of rapidity in
Au+Au reactions at $\sqrt s =200$~AGeV, min.bias.
\label{v1yp}}
\end{figure}

\begin{figure}[t]
\vskip 0mm
\vspace{1.0cm}
\centerline{\psfig{figure=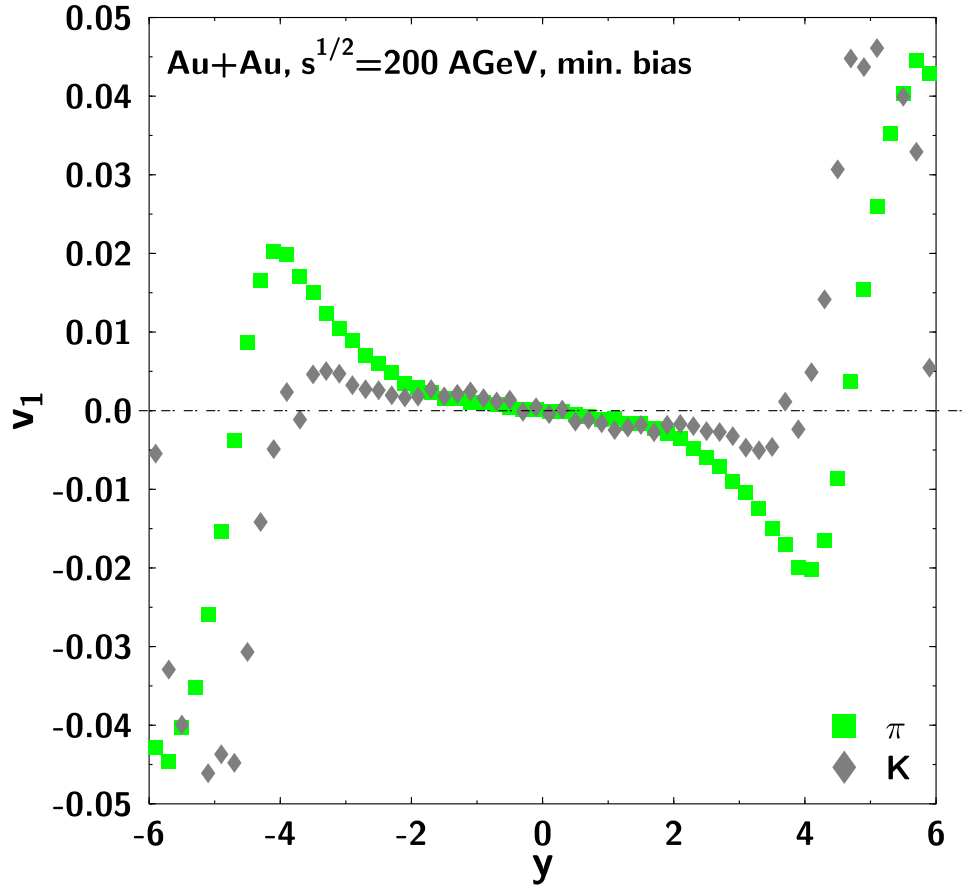,width=15cm,clip=}}
\vskip 2mm
\caption{Directed flow parameter $v_1$ of kaons and pions as a 
function of rapidity in Au+Au reactions at $\sqrt s =200$~AGeV, min.bias.
\label{v1ypi}}
\end{figure}

These arguments firmly establish the prediction of a
change of sign of the directed flow at mid-rapidity. 
The scenario is tested  quantitatively in Fig. \ref{v1yp} 
and \ref{v1ypi}  for Au+Au collisions at $\sqrt s$~=~200~$A$GeV at 
minimum trigger bias,
i.e. integrated over all impact parameters, using the 
Ultra-Relativistic Quantum Molecular
Dynamics (UrQMD 1.2) model in cascade mode~\cite{urqmd}. 
To quantify directed flow, the first Fourier
coefficient~\cite{posk98,volosz}, $v_1$, of the 
particle azimuthal distribution is used.
At a given rapidity and transverse
momentum the coefficient is determined by \cite{volosz}: 
\be
v_1 = \left\langle \frac{p_x}{p_t} \right\rangle \quad.
\ee
Similarly the second Fourier coefficient is determined from \cite{volosz}: 
\be
v_2=\left\langle \frac{p_x^2}{p_t^2}  
- \frac{p_y^2}{p_t^2} \right\rangle\quad,
\ee
which will be investigated later in this paper.

Figure~\ref{v1yp} shows the
UrQMD calculations of $v_1$ for nucleons, lambdas and anti-protons 
in Au+Au collisions at the full RHIC energy ($\sqrt s=200$~AGeV). 
Indeed, the shape of $v_1(y)$ is for nucleons and lambdas at mid-rapidity
consistent with the picture described above. Anti-protons show a
strong anti-correlation with the protons at forward/backward rapidities. 
This indicates the presence of anti-baryon absorption in nuclear matter even
at ultra-relativistic energies.
Zooming into the midrapidity region (Fig. \ref{v1y_22}): $v_1$
shows a weak negative slope at mid-rapidity for protons. The distribution of
$\Lambda$'s is in shape and magnitude similar to those of the protons.
For larger rapidities,
the $v_1$ values show the well-known 'bounce-off'
\cite{hofmann76a,stoecker86a,Stocker:1979mj}  of the 
nucleons around projectile and target rapidity. 
Note that the recently predicted wiggle in hydrodynamical calculations
\cite{brachmann,csernai} has a very different source: That 
wiggle occurs only at small impact parameters and if a phase transition to a
QGP is included. 
The QGP equation of state is a prerequisite to reach the stopping
needed to create a tilted source and the stall of the $p_x$ flow. 
The predicted wiggle in this Letter does not assume a
QGP equation of state. Another difference is that our
predictions rely on partial transparency. 

Pions and kaons are produced particles and their 
space-rapidity correlation is different
from that of nucleons as shown in Fig.~\ref{v1ypi} and Fig.~\ref{v1y_22}.
Shadowing by nucleons at central rapidities 
might also lead to an observable signature in the pion and kaon directed flow.
However, due to the small number of participating nucleons in semi-peripheral
collisions, this signature does only show up near target and projectile
rapidities. It does not lead to a wiggle in the $v_1$ of pions at midrapidity. 
Compared to the baryons and to the pions, the kaons show only very weak flow
over the whole inspected rapidity range.
At very forward/backward rapidities, the pions and kaons seem to 
follow the nucleons:
The light mesons are 'boiled-off' the excited spectator matter, thus
following the nucleons bounce-off flow pattern closely.

\begin{figure}[t]
\vskip 0mm
\vspace{1.0cm}
\centerline{\psfig{figure=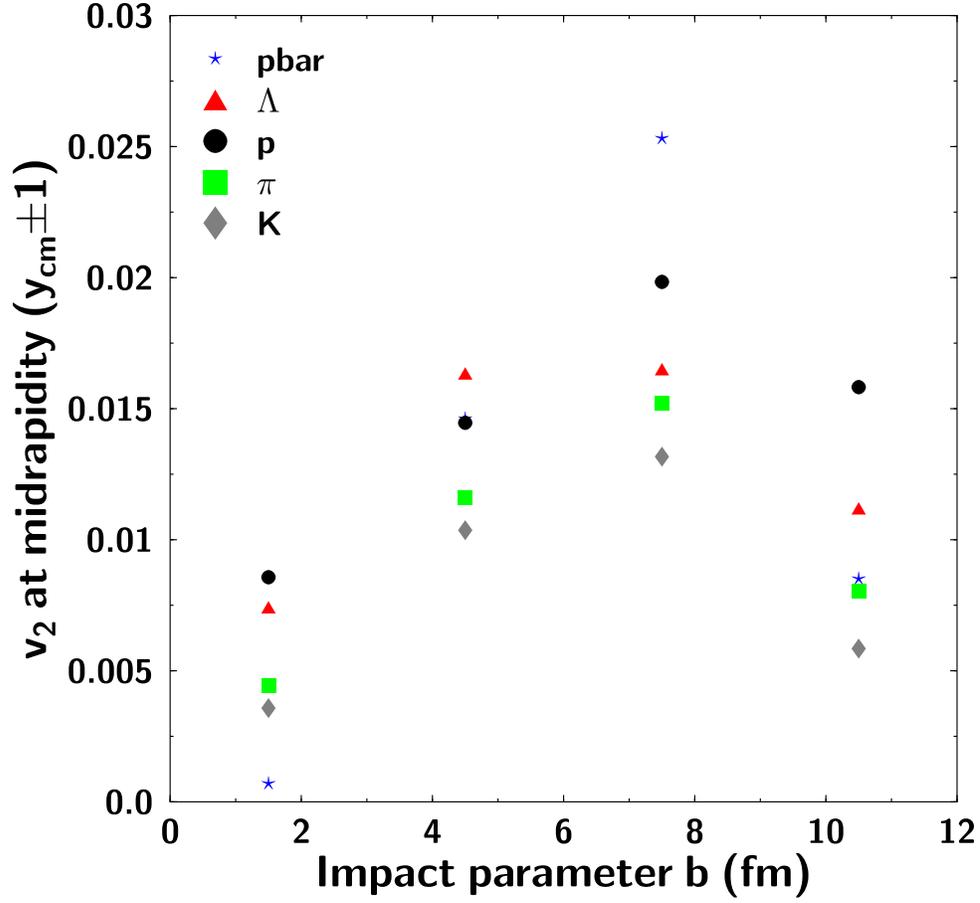,width=15cm,clip=}}
\vskip 2mm
\caption{Elliptic flow parameter $v_2$ of pions, kaons, protons, anti-protons
and lambda's at midrapidity ($|y|\leq 1$) as a function of 
centrality in Au+Au reactions at  $\sqrt s =200$~AGeV.
\label{v2bdep}}
\end{figure}

\begin{figure}[t]
\vskip 0mm
\vspace{1.0cm}
\centerline{\psfig{figure=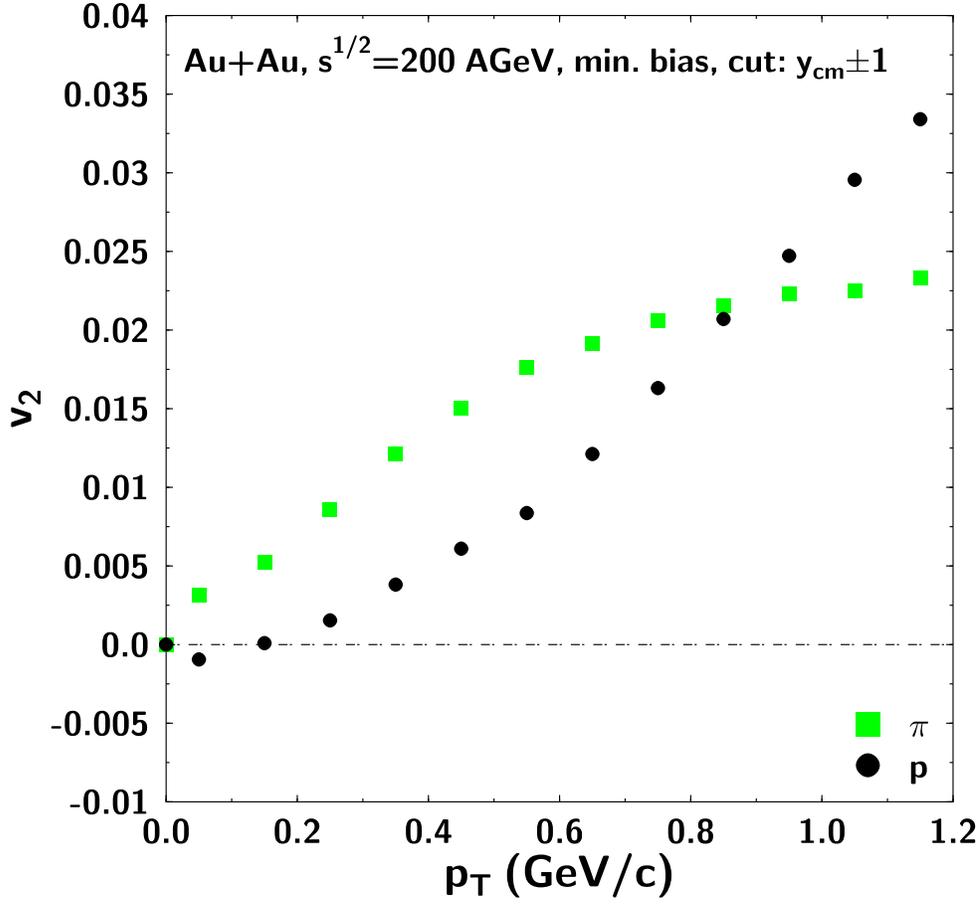,width=15cm,clip=}}
\vskip 2mm
\caption{Elliptic flow parameter $v_2$ at midrapidity ($|y|\leq 1$) as a 
function of transverse momentum 
in Au+Au reactions at $\sqrt s =200$~AGeV, min.bias.
\label{v2pt}}
\end{figure}

The elliptic flow of matter at central rapidities is another interesting
observable which yields crucial
information about the interaction strength and the mean free path of the matter
created at midrapidity. This information can be directly observed in the
strength of the $v_2$ parameters at midrapidity:
Fig. \ref{v2bdep} shows the impact parameter dependence of the elliptic flow
parameter for various particle species at midrapidity ($|y|\leq 1$). 
The impact parameter bins are: $b\leq 3$~fm, $3\leq b\leq 6$~fm, $6\leq b\leq
9$~fm and $9\leq b\leq 12$~fm.   
A clear maximum of the elliptic flow in all particle species 
is observed for semi-peripheral collision. 
Our speculation about the 
formation of transverse flow in those ultra-relativistic collision is also
supported by the transverse momentum dependence of $v_2$ (in min. bias Au+Au
reactions) as depicted in
Fig. \ref{v2pt}: A strong increase of the ellipticity parameter with $p_T$ 
is predicted which signals the existence of radial expansion.

\begin{figure}[t]
\vskip 0mm
\vspace{1.0cm}
\centerline{\psfig{figure=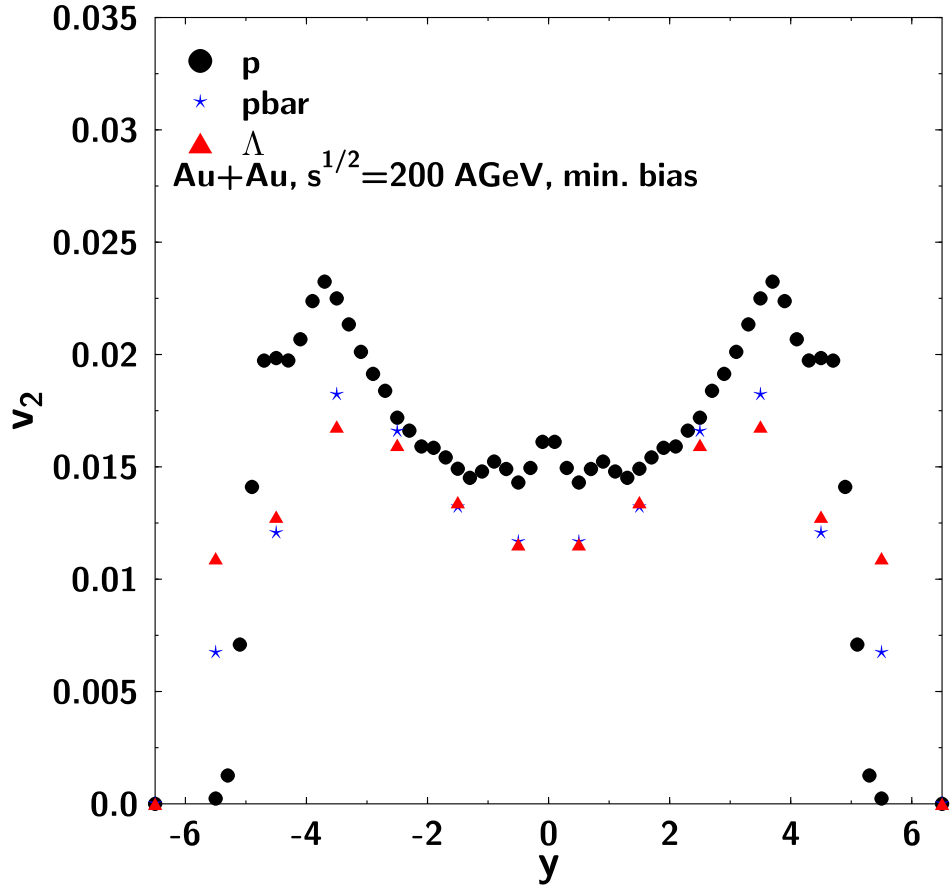,width=15cm,clip=}}
\vskip 2mm
\caption{Elliptic flow parameter $v_2$ of protons, lambdas and anti-protons 
as a function of rapidity in Au+Au reactions at $\sqrt s =200$~AGeV, min.bias.
\label{v2yp}}
\end{figure}

\begin{figure}[t]
\vskip 0mm
\vspace{1.0cm}
\centerline{\psfig{figure=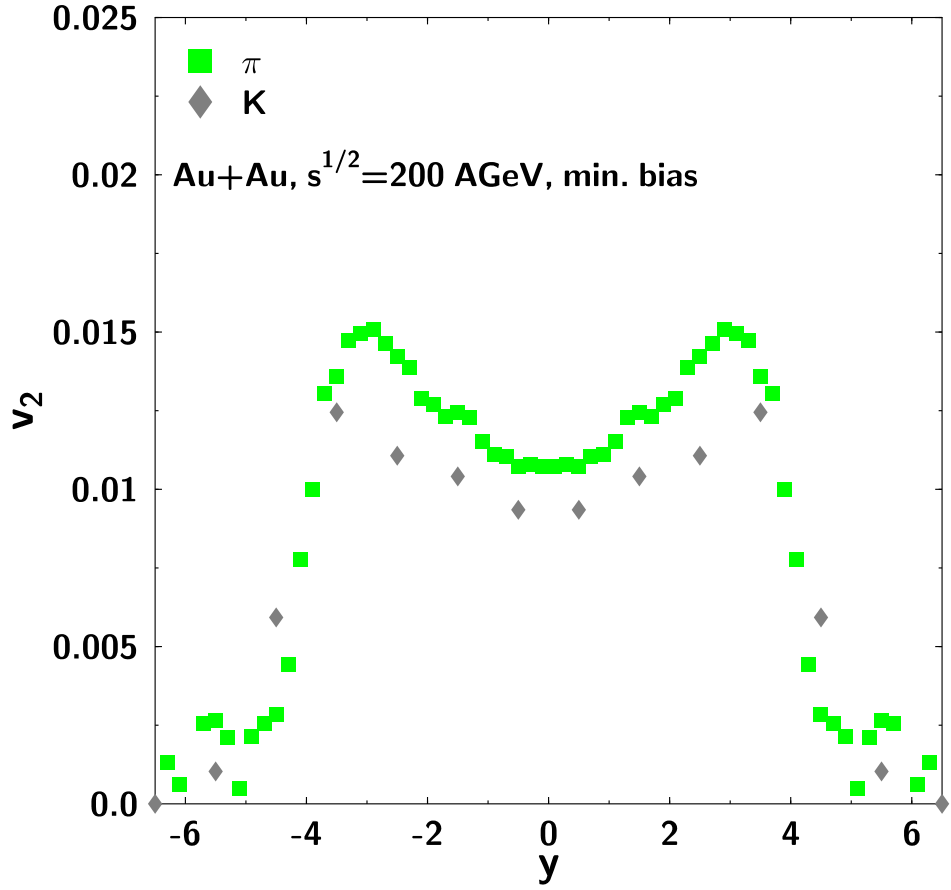,width=15cm,clip=}}
\vskip 2mm
\caption{Elliptic flow parameter $v_2$ of kaons and pions 
as a function of rapidity in Au+Au reactions at $\sqrt s =200$~AGeV, min.bias.
\label{v2ypi}}
\end{figure}

Figs. \ref{v2yp} and \ref{v2ypi} shows the anisotropic 
flow of nucleons, lambdas, anti-protons, pions and kaons  as a function of
rapidity for minimum biased Au+Au collisions at the full RHIC 
energy ($\sqrt s = 200$~AGeV). 
The elliptic flow of all inspected hadrons shows a prominent 
dip at central rapidities. This indicates a region of low `pressure' (or small
interaction strength, to be more specific). 
Comparing the elliptic flow parameters of baryons (Fig. \ref{v2yp}) 
with those of the mesons (Fig. \ref{v2ypi}) shows 
that the mesons acquire 2/3 of the
baryon elliptic flow. This scaling with the geometrical quark model cross
section may indicate a connection between elliptic flow and
in-medium cross section of the hadrons as will be discussed below.

\begin{figure}[t]
\vskip 0mm
\vspace{1.0cm}
\centerline{\psfig{figure=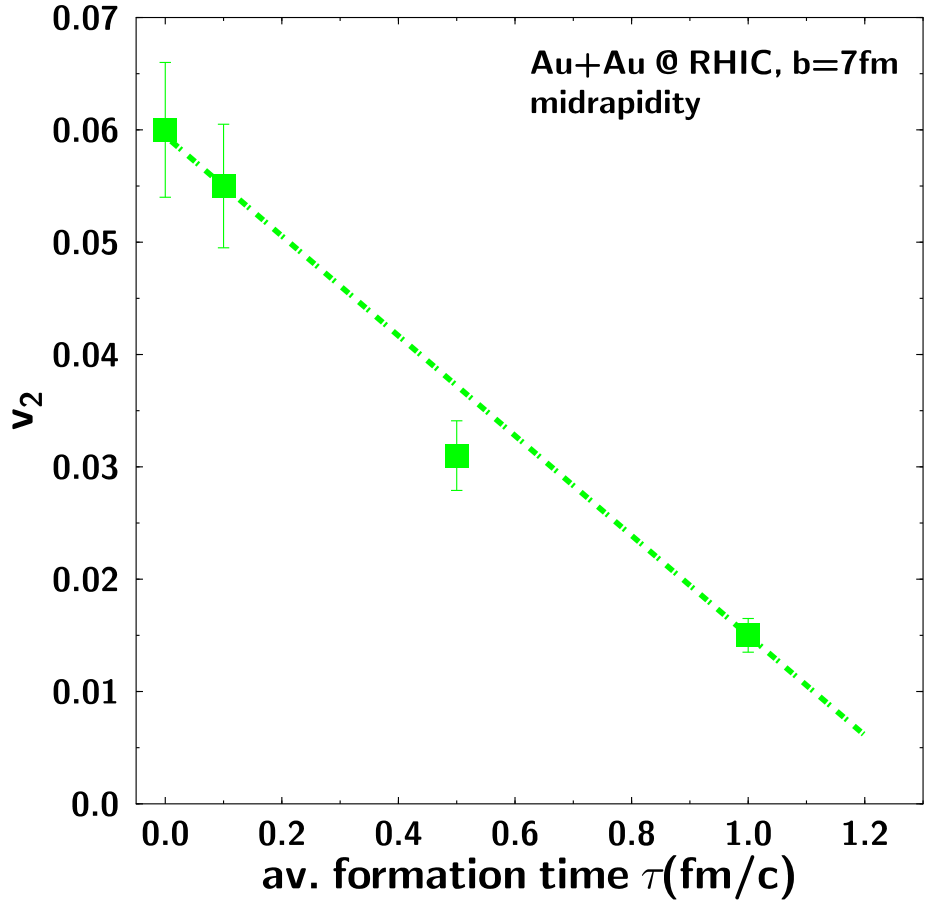,width=15cm,clip=}}
\vskip 2mm
\caption{Relation between the elliptic flow 
parameter $v_2$ at midrapidity and the mean free path (formation time) 
of the particles in Au+Au reactions at $\sqrt s =200$~AGeV, $b=7$~fm.
\label{v2f}}
\end{figure}

The appearance of this dip in $v_2(y)$ might appear to be 
counter-intuitive. However, it points
towards a distinct feature of the model dynamics in the early stage, 
namely the pre-equilibrium string dynamics and interactions on 
the parton level.
Fig. \ref{v2f} shows that the $v_2$ parameter is closely related to the
formation time of particles in the string picture. A standard default setting
of formation time results in an average formation time (in the string rest 
frame) of 1~fm/c. Consequently, the
particles created in the initial stage of the collision are not allowed to
interact during this time. With particle velocities near the speed of light,
these particles do have a mean free path on the order of 1~fm at midrapidity. 
If the formation time - and therefore the mean free path - is decreased, the
elliptic flow parameter $v_2$ increases. In the limit of a 
vanishing mean free path (hydro limit) the elliptic flow 
in the present model becomes maximal and it is quantitatively 
consistent with hydrodynamical predictions \cite{pasi}. 
In contrast, increasing the formation time (mean free path) results in a 
vanishing $v_2$, in line with the limit given by Hijing calculations without
quenching \cite{snelli}. 

Thus, the strength of the anisotropic flow of pions is 
directly connected to the mean free path of the particles (partons, hadrons)
forming the hot midrapidity  region. The measurement of $v_2$ might 
therefore yield valuable information about the transport properties of
QCD-matter, like the interaction frequencies and the 
viscosity of the excited partonic and hadronic matter at RHIC energies.
Especially $\Omega$ baryons, with their small hadronic cross section, 
are supposed to measure QGP properties without any
additional disturbance due to the hadronic phase.

In this Letter we have shown that a combination of space-momentum 
correlations and radial expansion results in a wiggle in the 
rapidity dependence of  directed flow in high energy 
nucleus-nucleus collisions. 
Since the magnitude of the wiggle depends on the radial expansion 
strength, which in turn is given by the equation of state, its observation 
provides unique insight into the
properties of the excited matter created around midrapidity. A peak in the
centrality dependence of $v_2$ is observed, while
a dip in the rapidity dependence of the elliptic flow
parameter is predicted. It has been demonstrated that the strength of the 
elliptic flow parameter is directly proportional to the characteristic 
mean free path of the matter formed in the central rapidity region.
Both the wiggle and the minimum in the elliptic flow appear at 
central rapidities, hence they are accessible to the STAR experiment 
at RHIC in the near future. 

\section*{Acknowledgements}

This research used resources of the
National Energy Research Scientific Computing Center (NERSC).
M. Bleicher wants to thank the A. v. Humboldt foundation, the Institut f\"ur
Theoretische Physik and the Nuclear Theory Group at LBNL for financial 
support. Fruitful discussions with Drs. N. Xu, R. Snellings and P. Huovinen are
gratefully acknowledged.

\newpage

\newpage

\newpage

\newpage

\newpage

\newpage

\newpage

\newpage

\end{document}